\documentclass[pdflatex,sn-mathphys-num]{sn-jnl}% Math and Physical Sciences Numbered Reference Style
\usepackage{graphicx}%
\usepackage{multirow}%
\usepackage{amsmath,amssymb,amsfonts}%
\usepackage{amsthm}%
\usepackage{mathrsfs}%
\usepackage[title]{appendix}%
\usepackage{xcolor}%
\usepackage{textcomp}%
\usepackage{manyfoot}%
\usepackage{booktabs}%
\usepackage{algorithm}%
\usepackage{algorithmicx}%
\usepackage{algpseudocode}%
\usepackage{listings}%
\usepackage{caption}
\usepackage[mathlines]{lineno} % mathlines 可让公式也编号
\usepackage{comment}

\theoremstyle{thmstyleone}%
\theoremstyle{thmstyletwo}%

\theoremstyle{thmstylethree}%

\begin{document}
%\linenumbers
\title[Article Title]{Defining the urban ``local'' with low dimensional manifolds of human mobility networks}


\author[1]{\fnm{Hezhishi} \sur{Jiang}}\email{jhzs@stu.pku.edu.cn}
\equalcont{These authors contributed equally to this work.}

\author*[2]{\fnm{Liyan} \sur{Xu}}\email{xuliyan@pku.edu.cn}
\equalcont{These authors contributed equally to this work.}

\author[2]{\fnm{Tianshu} \sur{Li}}\email{2301213451@pku.edu.cn}
\equalcont{These authors contributed equally to this work.}

\author[3]{\fnm{Jintong} \sur{Tang}}\email{tangjintong@pku.edu.cn}

\author[4]{\fnm{Zekun} \sur{Chen}}\email{chenzekun@pku.edu.cn}

\author[2]{\fnm{Yuxuan} \sur{Wang}}\email{2201213481@stu.pku.edu.cn}

\author[2]{\fnm{Haoran} \sur{Liu}}\email{2401213545@stu.pku.edu.cn}

\author*[5]{\fnm{Hongmou} \sur{Zhang}}\email{zhanghongmou@pku.edu.cn}

\author*[6]{\fnm{Huanfa} \sur{Chen}}\email{huanfa.chen@ucl.ac.uk}

\author*[3]{\fnm{Yu} \sur{Liu}}\email{liuyu@urban.pku.edu.cn}

\affil[1]{\orgdiv{Academy for Advanced Interdisciplinary Studies}, \orgname{Peking University}, \orgaddress{\street{5 Yiheyuan Road}, \city{Beijing}, \postcode{100871}, \state{Beijing}, \country{China}}}

\affil*[2]{\orgdiv{College of Architecture and Landscape Architecture}, \orgname{Peking University}, \orgaddress{\street{5 Yiheyuan Road}, \city{Beijing}, \postcode{100871}, \state{Beijing}, \country{China}}}

\affil*[3]{\orgdiv{School of Earth and Space Sciences}, \orgname{Peking University}, \orgaddress{\street{5 Yiheyuan Road}, \city{Beijing}, \postcode{100871}, \state{Beijing}, \country{China}}}

\affil[4]{\orgdiv{School of Mathematical Sciences}, \orgname{Peking University}, \orgaddress{\street{5 Yiheyuan Road}, \city{Beijing}, \postcode{100871}, \state{Beijing}, \country{China}}}

\affil[5]{\orgdiv{School of Government}, \orgname{Peking University}, \orgaddress{\street{5 Yiheyuan Road}, \city{Beijing}, \postcode{100871}, \state{Beijing}, \country{China}}}

\affil[6]{\orgdiv{Centre for Advanced Spatial Analysis}, \orgname{University College London}, \orgaddress{\street{Gower Street}, \city{London}, \postcode{WC1E6BT}, \country{the United Kindom}}}


\abstract{Urban science has largely relied on universal models\cite{Verbavatz2020,Schlapfer2021,Simini2012,Song2010,Pappalardo2015}, rendering the heterogeneous and locally specific nature of cities\cite{lefebrve1991production,tuan1979space,castells2011power} effectively invisible. Here we introduce a topological framework that defines and detects localities in human mobility networks. We empirically demonstrate that these human mobility network localities are rigorous geometric entities that map directly to geographic localities, revealing that human mobility networks lie on manifolds of dimension $\leq 5$. This representation provides a compact theoretical foundation for spatial embedding\cite{Jankowski2023,Gu2021} and enables efficient applications to facility location and propagation modeling. Our approach reconciles local heterogeneity with universal representation, offering a new pathway toward a more comprehensive urban science\cite{Lobo2020}.}

\keywords{Human mobility networks, locality, low-dimensional manifolds, spatial analysis, urban science}



\maketitle

\section{Introduction}\label{sec1}

The enduring tension between the local and the global lies at the heart of the social sciences\cite{lefebrve1991production,castells2011power,tuan1979space}. The local refers to the lived, everyday spaces shaped by people's practices, while the global denotes the broader forces and structures that organize and influence those spaces. This tension is also pervasive in urban science\cite{Lobo2020}, where prevailing paradigms often rely on global narratives to explain and predict complex urban phenomena. Such narratives employ universal equations or uniform mechanisms, for example, using a stochastic equation to model the dynamics of urban scales\cite{Verbavatz2020}, a single function to describe the visitation frequency of human mobility\cite{Schlapfer2021}, or an ideal gas model to simulate cross-scale geographic space from buildings to nations\cite{Boucherie2025}. Yet, as revealed by sociological studies on modernity and postmodernity, the local frequently exhibits resistance against the global forces\cite{lefebrve1991production}, leading to the failure of homogenized interpretations and plans. A central paradox is hence presented: the heterogeneous local, which is intuitively and empirically central to social life, remains invisible in the universal laws of urban science. Resolving this tension requires urban science to define "locality" in a way that accommodates its heterogeneous essence, enabling it to exist and be detected independently of its relation to the global.
%The enduring tension between the local and the global lies at the heart of the social sciences\cite{lefebrve1991production}. This tension is widespread in studies of human mobility and, more broadly, in urban science, where prevailing paradigms often rely on global narratives to explain and predict complex urban phenomena. Such narratives employ universal equations or uniform mechanisms, for example, using a stochastic equation to model the dynamics of urban scales\cite{Verbavatz2020}, a single function to describe the visitation frequency of human mobility\cite{Schlapfer2021}, or an ideal gas model to simulate cross-scale geographic space from buildings to nations\cite{Boucherie2025}. Yet, as revealed by sociological studies on modernity and postmodernity, the local frequently exhibit resistance against the global forces\cite{lefebrve1991production}, leading to the failure of homogenized interpretations and plans. A central paradox is hence presented: the heterogeneous local—which is intuitively and empirically central to social life—remains invisible in the universal laws of human mobility and urban science. Resolving this tension requires the study of human mobility and urban science to define "locality" in a way that accommodates its heterogeneous essence, enabling it to exist and be detected independently of its relation to the global.
%社科文献，支持local叙事的，还请许老师帮忙补充一些，哈维和卡斯特尔可以引，《认同的力量》（The Power of Identity）段义孚 space and place，空间与地方。德·赛托，查一下，走路使用与附近是什么书。吉登斯The Consequences of Modernity备选，可能放在讨论里，提供了这种分析的一种工具。

The inability to adequately define locality poses significant limitations for a comprehensive urban science, ironically hindering its pursuit of universal laws. This limitation is particularly evident in fields such as spatial network\cite{Barthelemy2011} and human mobility research. For instance, while the notion that network connections are governed by a spatial constraint is intuitive\cite{Barrat2005}, a universal functional form for distance decay, specifically the probability of forming an edge between two nodes decreases as the distance between them increases, has yet to be found. This lack of a universal equation extends to the classic debate between the gravity model\cite{Wilson1967}, the intervening opportunities model, and the radiation model\cite{Simini2012}, where there remains no consensus on which model is superior or what their optimal parameters and functional forms should be\cite{Mazzoli2019,Schlapfer2021,Boucherie2025,Simini2012}. Mathematically, this difficulty can be attributed to the inherent heterogeneity of local metrics, given that distance is an integral of such measures. The assumption of a universal distance decay function, by contrast, presupposes a homogeneous metric. We argue that a topological approach, inherently free from specific distance assumptions, provides a novel framework that fundamentally addresses this limitation.
%The inability to adequately define locality poses significant limitations for a comprehensive study of human mobility and urban science, ironically hindering its pursuit of universal laws. This limitation is particularly evident in fields such as spatial network\cite{Barthelemy2011} and human mobility simulation. For instance, while the notion that network connections are governed by a spatial constraint is intuitive\cite{Barrat2005}, a universal functional form for distance decay has yet to be found. This lack of a universal equation extends to the classic debate between the gravity model\cite{Wilson1967}, the intervening opportunities model, and the radiation model\cite{Simini2012}, where there remains no consensus on which model is superior or what their optimal parameters and functional forms should be\cite{Mazzoli2019,Schlapfer2021,Boucherie2025,Simini2012}. Mathematically, this difficulty can be attributed to the inherent heterogeneity of local metrics, given that distance is an integral of such measures. The assumption of a universal distance decay function, by contrast, presupposes a homogeneous metric. We argue that a topological approach, which by definition does not rely on specific distance assumptions, can effectively circumvent this limitation.

Beyond the paradigm of distance decay, existing urban complex systems research also employs a simulation approach\cite{Song2010,Pappalardo2015,Boucherie2025}. However, these models, built on minimum assumptions, remain constrained by uniform functional forms and behavioral rules for local agents. This creates a fundamental contradiction: under such minimum assumptions, existing models cannot be infinitely updated to incorporate new behavioral mechanisms. For example, successive models for human mobility (e.g., EPR\cite{Song2010}, d-EPR\cite{Pappalardo2015}, and PEPR\cite{Schlapfer2021}) have pushed the minimum assumption framework to its conceptual limits. Consequently, following this line of inquiry, our understanding of human mobility is logically prevented from exhausting all behavioral rules. This limitation is largely attributable to the inherent heterogeneity of localities and individuals, which cannot be captured by the iterative application of a single function. We propose that the concept of manifolds provides a powerful representation for this local heterogeneity. We argue that this new perspective, by focusing on representation rather than designing ever more complex behavioral mechanisms, could reveal a more comprehensive set of laws.
%Beyond the top-down paradigm of distance decay, existing human mobility and urban complex system research also employs a bottom-up approach\cite{Song2010,Pappalardo2015,Boucherie2025}. However, these models, built on minimum assumptions, remain constrained by uniform functional forms and behavioral rules for local agents. This creates a fundamental contradiction: under such minimum assumptions, existing models cannot be infinitely updated to incorporate new behavioral mechanisms. For example, successive models for human mobility (e.g., EPR\cite{Song2010}, d-EPR\cite{Pappalardo2015}, and PEPR\cite{Schlapfer2021}) have pushed the minimum assumption framework to its conceptual limits. Consequently, following this line of inquiry, our understanding of human mobility is logically prevented from exhausting all behavioral rules. This limitation is largely attributable to the inherent heterogeneity of localities and individuals, which cannot be captured by the iterative application of a single function. We propose that the concept of manifolds provides a powerful representation for this local heterogeneity. We argue that this new perspective, by focusing on representation rather than designing ever more complex behavioral mechanisms, could reveal a more comprehensive set of laws.

Here, we take human mobility, an inherently spatial practice\cite{de1998practice}, as our object of study. We adopt a topological rather than a distance-based perspective to represent the human mobility network. Within this framework, we define locality to construct what we term the \textit{human mobility manifolds}. Our approach shows that a locality is not merely a philosophical concept but a rigorous geometric entity. This provides a quantitative foundation for linking spatial theory in the social sciences with empirical patterns of urban mobility.

Building on Tobler’s First Law of Geography, which states that everything is related to everything else but near things are more related than distant things, we find that human mobility networks not only follow this law but also exhibit clear regional discontinuities. These discontinuities reveal the two defining properties: (1) the existence of localities, and (2) their mappability to geographic space. Together, they provide a rigorous definition of the manifold underlying human mobility. This construction serves as both a theoretical foundation and a practical tool that simplifies spatial analyses and enables efficient, interpretable modeling of mobility flows. The empirical findings are verified through case studies in diverse regions and scales, including Shenzhen, a Japanese city, Greater Mexico City, England, and Italy.

\begin{figure}[H]
\centering
\includegraphics[width=0.9\textwidth]{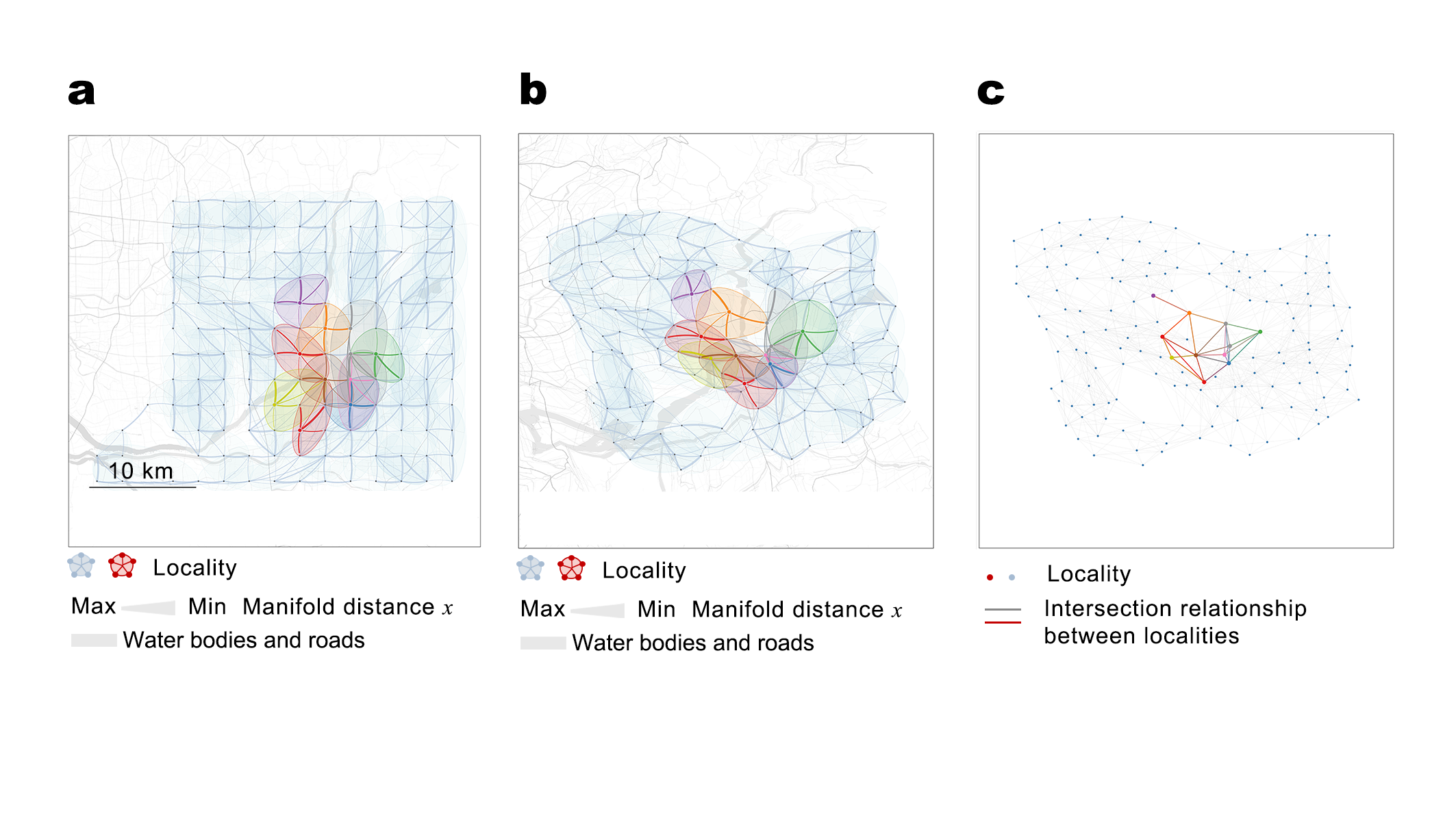}
\caption{\textbf{The locality of human mobility.} \textbf{a}, Localities of human mobility on a map of an undisclosed Japanese city. Irregular regions represent localities derived from normalized mobility flow data. Highlighted colored localities are examples from the pale blue regions, shown to trace correspondences between \textbf{a} and \textbf{b} and illustrate intersections. Each locality includes a central node, other nodes, and local edges connecting them. Edge thickness in both \textbf{a} and \textbf{b} is linearly normalized by manifold distance. \textbf{b}, Localities on the human mobility manifolds. Colors correspond to geographic counterparts in \textbf{a}, demonstrating that network localities map to geographic localities. To highlight correspondences, the road network and waterways are linearly mapped using right-angled triangles formed by the nodes, giving the appearance of a distorted map. \textbf{c}, Intersection network of localities. Nodes correspond to the same colored localities in \textbf{a} and \textbf{b}, now condensed to dots, and edges represent locality intersections. This network mirrors the intersection structures in both geographic and manifold spaces, showing that continuous percolation in 2D geographic space is equivalent to percolation on the mobility network, underpinning the manifold’s existence.}\label{fig0}
\end{figure}

\section{The two sufficient conditions}\label{sec2}

Mathematically, to define $d$-dimensional manifolds, we need to find a local open set for each point, and each open set is diffeomorphic to a $d$-dimensional Euclidean open set\cite{Guillemin1974}. Apart from these mathematical requirements, practically the majority of these open sets should path-connect with each other to make the embedding possible, which will be considered in a percolation way in this paper. Considering the discrete reality of human mobility  data, we will view the open sets as the locality defined in this work.

\begin{figure}[H]
\centering
\includegraphics[width=0.9\textwidth]{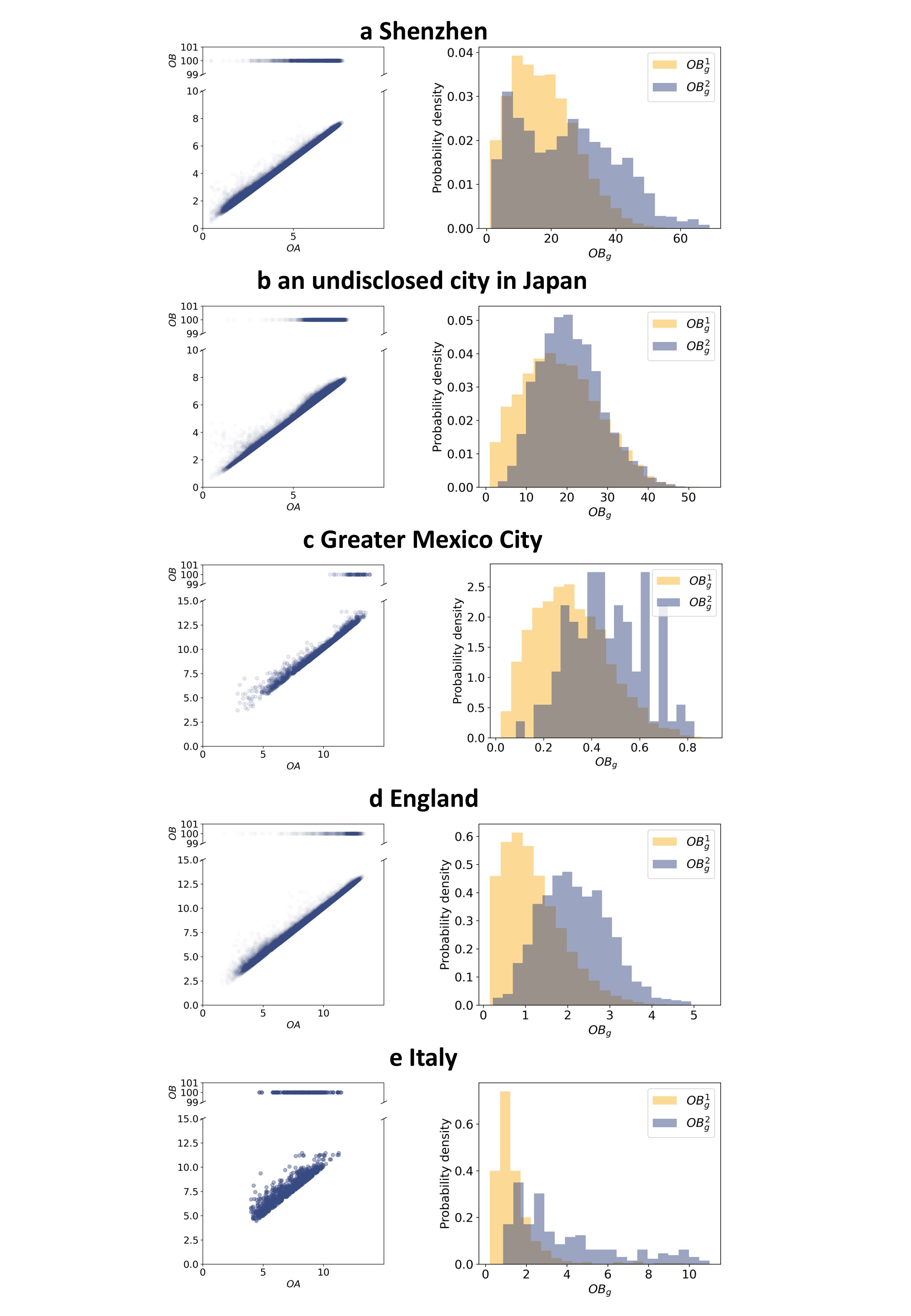}
\caption{\textbf{Empirical evidence for the two sufficient conditions guaranteeing the existence of human mobility manifolds.} The plots in the left column illustrate the dichotomy of $B$, serving as empirical evidence for the existence of locality. Points along the diagonal represent $B^1$, while those clustered along the upper horizontal line represent $B^2$. The lengths of ${OB}^2$ are all 100 because, for computational convenience, the distances between unconnected node pairs are artificially set to a sufficiently large number, which is 100 in our practice, to approximate what would theoretically be infinite distances. The plots in the right column provide empirical evidence for the second condition: when measuring the geographic distances between the two kinds of $B$ and $O$, we find ${{OB}^1}_g<{{OB}^2}_g$, namely those $B$ outside the network locality at $O$ are also geographically far away to $O$.}\label{fig1}
\end{figure}

As mentioned earlier, we pointed out two sufficient conditions for the existence of manifolds underlying human mobility networks (Fig. \ref{fig0}). The first is the existence of localities in human mobility networks, which is empirically verified through a dichotomy-based analysis. This analysis is inspired by the form of dichotomy of returners and explorers in human mobility\cite{Pappalardo2015}. Starting with any given centre node $O$, we explore its neighbors ($A$) by the order of increasing distances, namely decreasing edge weights, to see if the locality at $O$ can be expanded further by $A$’s locality. For $A$’s neighbors ($B$) that satisfies $OA<OB$ (notice that for $B$ with $OA \geq OB$, it is already explored before $A$), we observe that the $B$ closest to $O$ only has two possibilities (see the plots in the left column in Fig. \ref{fig1}): (1) Part of $B$s have $OA \approx OB$, which we denote as $B^1$; (2) The other part has $OA \ll OB$, which means $B$ is weakly or not connected to $O$, denoted as $B^2$. This dichotomy pattern shows the existence of localities: $B^1$ means the locality at $O$ can still expand by part of $A$’s locality; while $B^2$ means the boundary of $O$’s locality is between $B^2$ and $A$. Conversely, if the locality does not exist, then either the boundary shall not exist, causing the absence of $B^2$, or with ambiguous boundary, forming a dispersion of $B$ in the triangle area over the diagonal in left column plots of Fig.\ref{fig1}. In the mathematical definition of a manifold, local open sets need to be defined for each point. This requirement is satisfied in our dichotomy analysis by centring the procedure on every node in the network. The diffeomorphism from a local open set on the manifold to a local open set in Euclidean space is, in the context of discrete data and scenarios, simplified to an isometric embedding.

The second condition is the existence of the mapping from human mobility network localities to geographic localities, which can find evidence in the following three aspects: (a) empirically, we analyse the distribution of geographic distances for the two kinds of $B$ to $O$, denoted as ${OB}_g$, and we find that ${{OB}^1}_g$ is significantly smaller than ${{OB}^2}_g$ by t-test, with $p$-value less than 0.0001 for all the cases (see the plots in the right column in Fig.\ref{fig1}). (b) Tobler’s first law\cite{Tobler1970} reported similar facts, though in another form, i.e. near things are more related than distant things. (c) human mobility laws provide supporting evidence, including the distribution of step length and radius of gyration\cite{Gonzalez2008}, the container theory\cite{Alessandretti2020}, and the visitation law\cite{Schlapfer2021}. The correspondence between localities of the same color in panels \textbf{a} and \textbf{b} of Fig. \ref{fig0} illustrates an example of this mapping.

Given the two conditions, we can now prove the existence of low-dimensional manifolds behind human mobility networks. The whole picture is as follows. (i) human mobility networks have localities. (ii) We map the network localities to the geographic localities. (iii) As the geographic space is 2-dimensional, we can apply the continuum percolation thresholds for two dimensions, which is 5.6 at most (for common shapes)\cite{Mertens2012}. The percolation threshold is defined as the ratio of the total area covered by the shapes surrounding each point to the total area of the field. As the number of localities equals the number of nodes, this threshold corresponds to the size of each locality. When area is replaced by the number of nodes, the threshold instead reflects the number of nodes to be contained within each locality. (iv) Thus the 6-node localities will path-connect with each other in a percolation way in the geographic space. (v) the intersecting relationship of localities in the geographic space is preserved back in the network, as shown in Fig. \ref{fig0} \textbf{c}. (vi) the percolation that rely on the intersecting relationship is also preserved back in the network. (vii) 6-node localities give the maximal dimension of 5 (diffeomorphism is simplified to a local isometric embedding, due to the discrete nature of human mobility and its data). (viii) Thus 5-dimensional human mobility manifolds can be constructed (the triangular inequality will be guaranteed in the next section). This is a QEF (Quod Erat Faciendum) style proof, but it shows why and how human mobility is sufficiently low dimensional manifolds. 

\section{The embedding of human mobility manifolds}\label{sec3}
Previously, we formally proved the existence of human mobility manifolds. To verify this proposition, we need to embed real-world human mobility data into 5-dimensional manifolds using a simple method. Emphasizing the simplicity of the method is intended to highlight the usefulness of the proven conclusion itself, rather than the computational performance achieved through a complicated algorithmic design.

Before performing the embedding, we first need to define a distance function that maps the edge weights of the human mobility network to a distance on the manifold, and ensure that this function satisfies two requirements: (1) it must obey the triangular inequality; and (2) it must ensure consistency between topological embedding and metric embedding (we will define this issue in detail later). The first requirement is achieved by introducing the edge weight-distance function from the hyperbolic model for weighted networks\cite{Allard2017}, where triangular inequality is used to identify the coupling of network topology and weights (see Methods and equation \ref{eq1} for details).  The second requirement is met by generalizing equation \ref{eq1} to maintain a global consistent radius of network localities (see Methods for details). Note that the unified neighborhood radius used here is derived from normalization incorporating local heterogeneity, and is thus not a predetermined radius.

\begin{figure}[H]
\centering
\includegraphics[width=0.9\textwidth]{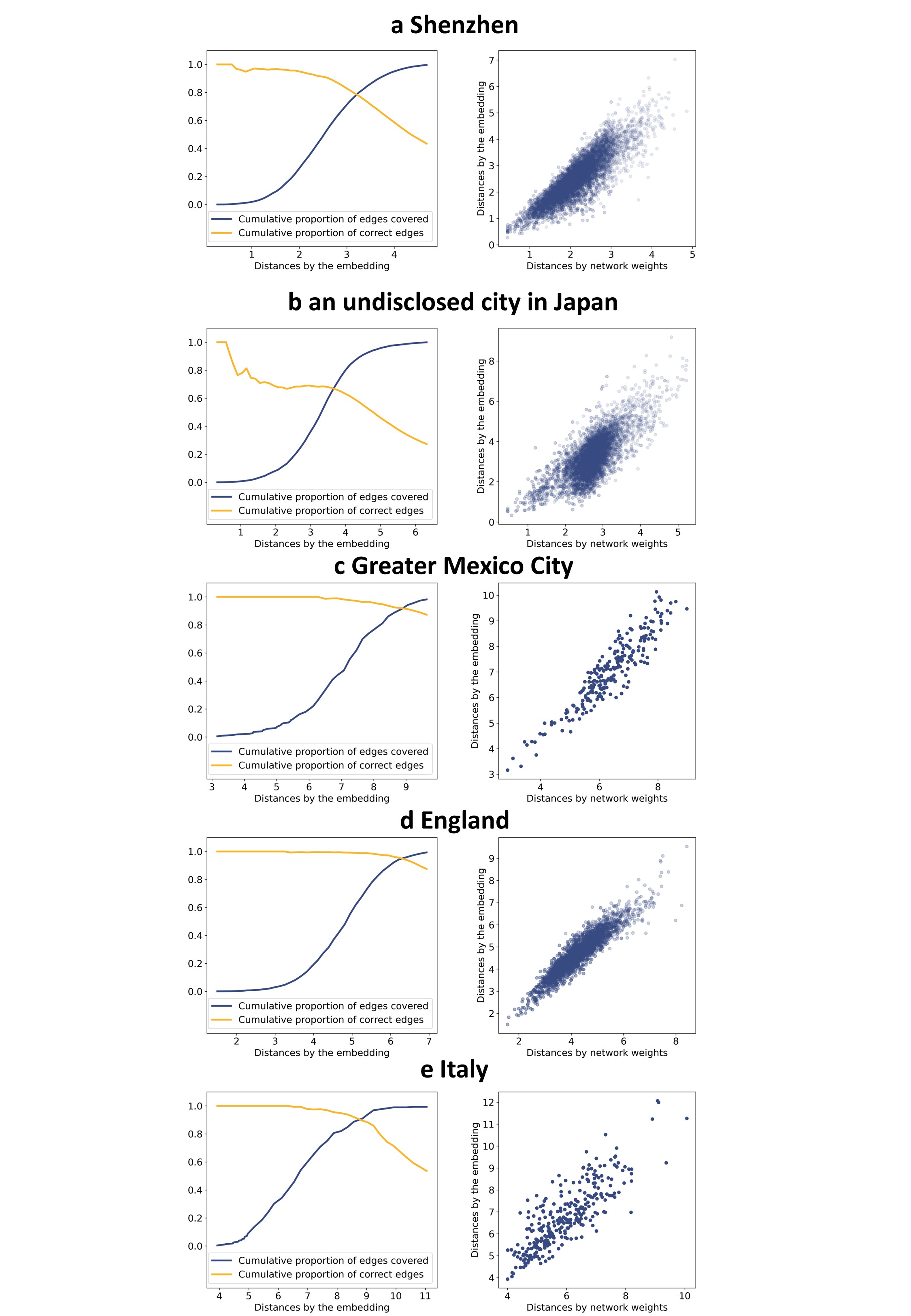}
\caption{\textbf{The embedding quality of human mobility manifolds, measured both topologically and metrically.}  The plots in the left column represent the topological embedding quality. We select node pairs whose embedding distances fall within an increasingly large distance window (with window size as $x$-axis) and examine two metrics: the proportion of these pairs that are actual local edges in the original graph (orange curve), and the proportion of all local edges that are covered by these selected node pairs (blue curve). A higher intersection point of the two curves indicates better topological embedding quality. The heights of the intersections are 0.78, 0.69, 0.92, 0.81, and 0.90 for subfigures \textbf{a–e}, respectively. The plots in the right column represent the metric embedding quality. The Pearson correlation coefficients for metric embedding in subfigures \textbf{a–e} are, 0.87, 0.75, 0.92, 0.98, and 0.81, respectively.}\label{fig2}
\end{figure}

After defining the distance function, we use TCIE\cite{Rosman2010} (Topologically Constrained Isometric Embedding) to perform the embedding. This algorithm is a variant of the well-known Isomap\cite{Tenenbaum2000} and addresses its failure on non-convex manifolds\cite{Rosman2010}. The failure arises because shortest paths crossing manifold boundaries are distorted and do not represent true geodesics. TCIE resolves this by detecting boundaries and excluding node pairs that intersect or lie too close to them from the optimization. The results demonstrate excellent topological and metric embedding performance (see Fig. \ref{fig2}). Since the algorithm is relatively intuitive (especially when compared with black box methods like graph neural networks, and UMAP\cite{McInnes2018}), the quality of the embedding can be regarded not as an artifact of algorithmic design, but as evidence supporting the validity of the theoretical conclusion established earlier.

\section{Spatial analysis on the human mobility manifolds}\label{sec4}

Since we have proved the existence of human mobility manifolds, what role do they play in spatial analysis that depends on human mobility networks? In this section, we illustrate through two scenarios, inlcuding location choice and propagation, where the application of manifold can greatly simplify the analysis and lead to succinct patterns. This suggests that human mobility manifolds have the potential to transform optimisation and simulation problems into geometric analysis. It is important to note that although we previously proposed a reliable embedding method, it is not the only possible one. Therefore, in this section, we adopt embedding methods that are more suited to specific application scenarios. In other words, we rely on the conclusion that human mobility manifolds exist, rather than the specific way to define the embedding. In fact, for any site or scenario, as long as the two conditions for the existence of a manifold can be verified (which is highly likely, given our extensive empirical work and supporting evidence from previous studies), and the distance function satisfies the triangular inequality, the manifold can be properly defined. The dimensionality of the manifold can be chosen based on embedding accuracy and practical considerations, within the constraint of the previously established upper bounds of five dimensions. For the sake of visual intuition, we chose a two-dimensional representation for spatial analysis, which is justified by the success of application.
%这段尝试简化了一下，原本有点啰嗦：Having established the existence of human mobility manifolds, we next explore their role in spatial analyses based on mobility networks. We illustrate two scenarios—location choice and propagation—showing that manifolds can simplify analysis and reveal concise patterns. This suggests that human mobility manifolds can transform optimization and simulation problems into geometric ones. While we previously proposed a reliable embedding method, it is not the only option. Here, we adopt embeddings suited to specific applications, based on the manifold’s theoretical existence: a valid manifold can be defined whenever the existence conditions are satisfied (as supported by empirical evidence) and the distance function obeys the triangle inequality. Its dimensionality can be chosen based on embedding accuracy and practical considerations, with established upper bounds of five dimensions or fewer. For visualization, we adopt a two-dimensional representation, justified by successful applications.

\begin{figure}[H]
\centering
\includegraphics[width=0.9\textwidth]{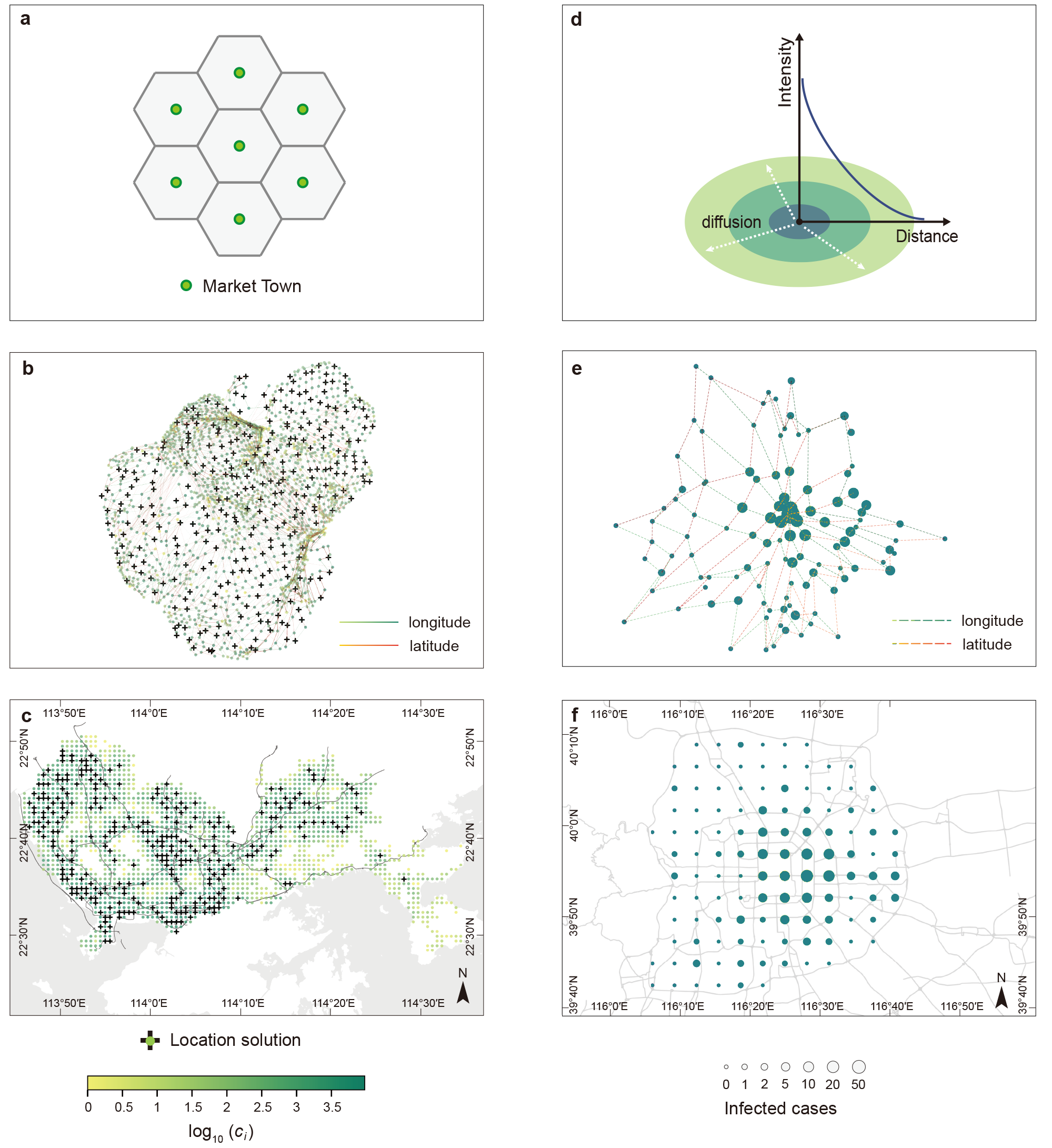}
\caption{\textbf{The location choice problem and the propagation problem on the human mobility manifolds.} \textbf{a-c}, The location choice problem (the case of Shenzhen). The node colors in \textbf{b, c} depict $\log_{10} \|c_i\|$, where $c_i$ is the set of individuals covered by the grid (namely node), see Methods for the definition of coverage; the cross marks depict optimal location solutions. The classic location choice law under the featureless plain assumption\cite{Christaller1933} gives rise to uniformly laid location patterns \textbf{a}, which are achieved with real-world data in \textbf{b}. The uniformity of optimal location solutions on the human mobility manifolds is evident both quantitatively (see Methods) and when compared with that on the geographic space \textbf{c}. \textbf{d-f}, the propagation problem (the case of Beijing). The node size in \textbf{e, f} depicts the number of infected cases in that cell. The classic propagation law would predict a concentric diffusion pattern in the featureless space \textbf{d}, which once again, is achieved with real-world data in \textbf{e}. The concentric test is described in the Methods, and is also evident when compared with the propagation on the geographic space \textbf{f}.}\label{fig3}
\end{figure}

The relation of location choice and human mobility started even before the emergence of human mobility data, when economists developed the flow capturing problem using spatial interaction models\cite{Hodgson1990}, aiming to optimize facility locations to capture as many simulated human flows as possible. With the advent of human mobility data\cite{Batty2012}, modelled flows based on assumptions have been replaced by real trajectory data, yet solving these location choice problems often relies on complicated optimisation\cite{CHURCH1974}. Inspired by central place theory\cite{Christaller1933}, that on a featureless plain, where demand is uniformly distributed and the serving of demand depends solely on distance, the ideal facility layout would be uniform (Fig. \ref{fig3}a). Can human mobility manifolds, through appropriate population equalization, serve as this featureless plain, thereby leading to an evenly distributed optimal facility (i.e. location solution) layout? The answer is yes. Using Shenzhen as a case study, we construct a manifold based on a distance function defined by the overlap ratio of users covered by pairs of locations/nodes. We then apply a cartogram\cite{Gastner2004} adjustment to the manifold for population equalization, and find that the optimal facility layout exhibits a uniform pattern (Fig. \ref{fig3}b; see Methods for details on manifold construction and the uniformity test). Apart from the quantitative uniformity test, the comparison between the optimal facility layout on the human mobility manifold and those on the geographic space (Fig. \ref{fig3}c) clearly reveals this uniformity. Note that we did not perform location choice optimization directly on the manifold; rather, we observed that the optimal solutions of conventional coverage-based location choice exhibit a uniform pattern when projected onto the manifold.

Human mobility data has also empowered the propagation modelling, which is typically achieved using complicated agent-based models\cite{Aleta2022} or differential equations\cite{Venkatramanan2021}. Our approach to simplifying the modelling of propagation is also inspired by classical theory, similar to the location choice scenario. Specifically, on an idealized featureless space, concentric diffusion patterns have long been predicted\cite{Hagerstrand1952} (Fig. \ref{fig3}d). It turns out that human mobility manifolds once again established this featureless plain. We analysed an empirical case of a COVID-19 outbreak in Beijing in June 2022 (see Methods for details), and found that on the human mobility manifold, the spatial distribution of infection cases forms an isotropic concentric pattern (Fig. \ref{fig3}e). The isotropy is not only statistically significant (see Methods for detail), but also visually evident when compared with the propagation on the geographic space (Fig. \ref{fig3}f).

\section{Discussion}\label{sec5}

In urban science, the heterogeneous local and its tension with global narratives have been largely neglected, despite their profound influence in the social sciences\cite{lefebrve1991production,tuan1979space,castells2011power}. Our work addresses this gap by adopting a topological perspective. We use a simple dichotomy, to be more precise, whether a locality's expansion encounters a boundary, to empirically demonstrate the ubiquitous existence of localities within human mobility across diverse regions and scales. Due to the very nature of topological analysis, the localities defined here do not change with the spatial scope of inquiry, nor do they need to follow the rules of a global narrative. Our analysis offers urban social sciences an effective instrument for examining the tensions between the global and the local\cite{giddens1985consequences,lefebrve1991production}.
%In the study of human mobility, and more broadly in urban science, the heterogeneous local and its tension with global narratives have been largely neglected

Our topological definition of localities addresses a fundamental theoretical gap: although metric-based and geometric methods are widely applied\cite{Taylor2015,Brockmann2013,Boguna2021}, their underlying structures have been only empirically validated. Because empirical evidence can never be exhaustive, it has remained unclear whether errors in these methods arise from model limitations or from flawed geometric assumptions. By leveraging two key properties of the localities defined here, we provide a proof of the human mobility manifold’s existence, mitigating this theoretical risk. In doing so, we bridge long-standing traditions in geography and network science: geography represents Earth-surface processes within geographic space\cite{Tobler1970} or its deformations\cite{Tobler1961}, whereas network science embeds networks in hidden geometric spaces\cite{Dall2002}. The manifold framework unifies these perspectives and establishes a foundation for cross-disciplinary dialogue. Concretely, in both geographic and network science models, as long as a well-defined distance can be specified that induces a manifold, further exploration of the model can be reframed as a geometric investigation of that manifold. For example, the hyperbolic distance used in this study\cite{Allard2017} can be regarded as inverted from a generalized gravity model, providing a convenient link between our manifold-based approach and established modeling traditions.

From a practical perspective, human mobility manifolds reduce complex optimization and modeling problems to geometric pattern analysis, as illustrated by applications in location-choice optimization and propagation modeling. More broadly, the introduction of manifolds offers an answer to a central question in spatial analytics: what constitutes an informative spatial representation and which of its elements deserve attention. This question that has gained increasing recognition in recent advances\cite{Bronstein2017,Zheng2023}.

\section{Methods}\label{sec11}
\subsection{Data description}\label{subsec1}
The empirical analysis in this study utilizes human mobility networks from five sites, defined by origin-destination (OD) flow data. To convert the original directed human mobility flows into an undirected network, we applied bidirectional summation to the OD flows. The undirected property is an intrinsic requirement of metric and geometric analysis, as by definition any metric should have commutativity. The selection of these sites takes into account the diversity of coverage areas and spatial scales. The data for Shenzhen comes from anonymized and gridded mobile location data (namely individual trajectories) from the Baidu app, while the data for an undisclosed city in Japan, Greater Mexico City, England, and Italy is sourced from open-access flow datasets\cite{Zhong2024,Yabe2024,Flores-Garrido2024,Pepe2020}.

The data for Shenzhen is anonymously collected from location services provided by Baidu. Given Baidu's extensive market share in China, the dataset covers a large user base, and our dataset records trajectories of 823,975 individuals. The data spans the entire day of 4 December, 2019, with a spatial resolution of 150-meter grids. We performed data cleaning and stay-point extraction following a grid-based trajectory analysis method\cite{Vazifeh2019}, and subsequently computed the flows. To improve computational efficiency, we aggregated the 150-meter grids in a 6×6 way, resulting in a total of 1,916 nodes.

The trajectory data for the anonymous city in Japan covers an undisclosed 75-day period\cite{Yabe2024} and is based on mobile location data provided by Yahoo, with 100,000 individuals. The data has a spatial resolution of 500 meters and a temporal resolution of 30 minutes. We directly extracted flow information from the trajectory data and performed a 5×5 grid aggregation, resulting in a total of 1,598 network nodes.

The data for Greater Mexico City comes from a flow network at the municipality level on 1 January, 2020\cite{Flores-Garrido2024}. This dataset is based on mobile location data provided by Veraset\cite{Flores-Garrido2024}, covering 33 million devices. In our network, there are a total of 76 nodes.

The open-source flow data for England is from November 2021 and covers 1\% of the population\cite{Zhong2024}. The spatial units are based on the H3 geospatial indexing system (https://h3geo.org/). The original data is in hexagonal cells of 0.1 square kilometers, which we aggregated upward into larger hexagons of 252 square kilometers to improve computational efficiency, and the number of nodes in our network is 794. A finer-grained analysis for England and London is provided in Extended Data Figs. \ref{edfig0} and \ref{edfig0-1}.

The data for Italy consists of flow information provided by Cuebiq Inc., based on 170,000 individuals, with spatial resolution at the level of Italian provinces\cite{Pepe2020}. We used the flow data from 18 January, 2020, to construct the network, which includes 107 nodes.

For the validation of the propagation modelling, we use the spatial distribution of cases from a COVID-19 outbreak in Beijing in June 2022, with case records from the official website of the Beijing Municipal Health Commission\cite{BeijingMunicipalHealthCommission2022}. As of 22 June, when the outbreak ended, there were a total of 388 infected cases, of which 361 were first-round cases. Considering that Beijing initiated emergency epidemic prevention measures after the first round of epidemic propagation (June 9), which may introduce bias in the modelling, this study only focuses on the first-round cases. After geocoding, there are 349 addresses in the first-round cases that could be used for further analysis, and the remaining 12 addresses are outside the study area (see Section 1 of Supplementary Information for the definition of the study area).

The mobile phone dataset used for the network construction of the propagation modelling is provided by one of the largest communication operators in China. The characteristics of this dataset are consistent with those described in a previous study\cite{Si2022}, where detailed data conditions can be found. (see Section 1 of Supplementary Information for the introduction of this data). 

\subsection{Details for the embedding of human mobility manifolds}\label{subsec2}

\subsubsection{Definition for the distance function}\label{subsubsec2}

To ensure that the distance function we define satisfies the triangle inequality, we adopt the distance function from the hyperbolic model for weighted networks\cite{Allard2017}. In this function, $w_{ij}$, and $x_{ij}$ denote the edge weight (i.e. the size of flow), and the hidden distance for the node pair $i$ and $j$. The distance function defined in this study is obtained by inversely solving the following equation for $x_{ij}$.

\begin{equation}
w_{ij}=\epsilon_{ij}\frac{\nu}{\mu^\alpha}\frac{\sigma_i \sigma_j}{\kappa_i \kappa_j}e^{-\frac{\alpha}{2}(x_{ij}-r)}\label{eq1}
\end{equation}

where $\sigma_i$ and $\kappa_i$ denote the hidden node strength (whose expectation equals the actual node strength) and the hidden degree (whose expectation equals the actual node degree) of node $i$, respectively. $\epsilon_{ij}$ denotes a random noise term for the node pair $i$ and $j$. $\nu$ and $\mu$ are normalization parameters. $\alpha$ is a trade-off parameter balancing the influence of hidden degrees and hidden distances on the network weights, and is determined to satisfy the triangular inequality. $r$ is the radius of the hyperbolic network model. By taking the inverse of Equation (1), we obtain a function that computes the distance based on the edge weight, node degree, and node strength in the real network:

\begin{equation}
x_{ij}=C+\ln{\frac{S_i S_j}{w_{ij}}}\label{eq2}
\end{equation}

where $C$ is a constant that consolidates various constants from Equation \ref{eq1}, and $S_i=\sigma_i/\kappa_i$. Since in Equation \ref{eq1}, the parameter $\alpha$ is determined by the triangular inequality, and changes of $\alpha$ correspond to changes in $C$ in Equation \ref{eq2}, the distance function will satisfy the triangle inequality through an appropriate choice of $C$. Specifically, we analyze all the triangles $ijk$ in the network and compute the difference between the longest edge (denoted as $x_{ij}$) and the sum of the two shorter edges ($x_{jk}+x_{ki}$). Since $x_{ij}-x_{jk}-x_{ki}=\ln{[(w_{jk}w_{ki})/ (w_{ij}{S_k}^2)}]-C$, to ensure that the triangle inequality is satisfied for the vast majority of triangles (operationally set to 99.99\%), we simply sort $\ln{[(w_{jk}w_{ki})/ (w_{ij}{S_k}^2)}]$ in descending order and take its 0.01\% quantile as one candidate of $C$. The choice to ensure that the vast majority of triangles, rather than all of them, satisfy the triangular inequality is made for the sake of stability (the very top of $\ln{[(w_{jk}w_{ki})/ (w_{ij}{S_k}^2)}]$ can vary a lot), and similar operations can be found in previous studies for network triangular inequalities\cite{Allard2017}. Note that the choice of $C$ should also ensure all the distances are nonnegative, which gives the other candidate of $C$, we use the larger one of the two as $C$. 

The second requirement for the distance function involves topological and metric embedding. Specifically, the topological embedding should ensure that node pairs that are close in the embedded space are actually connected and within locality, while node pairs that are far apart are either unconnected or not part of the same locality. This locality is defined, as in the theoretical proof section, using the k-nearest neighbours. Metric embedding is more straightforward—it requires that the embedded distances of the local edges have a monotonic one-to-one correspondence with the actual edge weights in the network. In practice, however, a good metric embedding does not always guarantee correct topology. This is because, if the distance function is poorly defined, the maximum distance from the centre node to the farthest node within the locality (namely the radius of the locality) can vary. Such variation makes it impossible to define a global threshold where node pairs within the threshold represent local topological connections, and those beyond it represent non-local or disconnected node pairs. In such cases, even if the embedding method achieves a good metric embedding, it cannot effectively capture the local topology.

It is worth noting that the form of Equation \ref{eq2} has the potential to unify topological embedding and metric embedding. By setting $S_i={{\sigma_i}^\eta} / {\kappa_i}^\xi$ and using the radii of all the localities as the basis, we estimate the parameters $\eta$ and $\xi$ through regression based on the relation $\ln{w_{ij}}=\ln{C_1 S_i S_j}=\ln{C_1}+\eta \ln{\sigma_i \sigma_j}-\xi\ln{\kappa_i \kappa_j}$. This ensures that the boundary threshold has an expected value of $C-C_1$, where $C_1$ is the intercept term in the regression. Moreover, since the regression coefficients alter the definition of distance, and consequently change the definition of locality and the radius, we iterate this process until the coefficients converge.

\subsubsection{A brief introduction to TCIE and principles for determining the parameters}\label{subsubsec2}

The TCIE algorithm\cite{Rosman2010} consists of three main steps: (1) boundary identification, (2) computation of the weight matrix for the objective function, and (3) optimization of the objective function.

In the boundary identification step, a voting strategy is employed\cite{Rosman2010}. Two approaches identifying boundary points by analyzing local neighborhoods around each node $i$, denoted as $\mathcal{N}(i)$, are applied, along with a linear dimensionality reduction for the local matrix of $\mathcal{N}(i)$. Based on our experience, a neighborhood size of 30 is appropriate. The first approach utilize the distance between the barycenter of $\mathcal{N}(i)$ and $i$ as an indicator of boundary likelihood. A larger distance suggests that $i$ is more likely to lie on a boundary. By specifying a parameter $n$ (based on our experience, 90\% of the total number of nodes in the network is proper), the top $n$ nodes with the largest such distances are selected as boundary candidates. The second approach is based on the directional structure within $\mathcal{N}(i)$’s local embedding. For a given node $j\ne i$ in $\mathcal{N}(i)$, a perpendicular to $ij$ at $i$ divides $\mathcal{N}(i)$ into two parts. The proportion $\pi_j^i$ is defined as the number of nodes $k$ forming an obtuse angle $\angle kij$, divided by those forming an acute or right angle. The number of $j$ with $\pi_j^i$ smaller than a threshold (operationally set to 0.9, the results have been tested to be insensitive to this parameter), namely $\|\{j|\pi_j^i<0.9\}\|$, indicates how node $i$ is likely to be a boundary node. The top $n$ nodes $i$ with the  largest $\|\{j|\pi_j^i<0.9\}\|$ are selected as another set of boundary candidates. Finally, the intersection of the candidate sets obtained from both approaches is taken as the final set of boundary points.

When calculating the weight matrix for the objective function, four cases need to be considered. The first case is when the shortest path between node pair $ij$ in the network does not intersect the boundary; in this case, the weight is set to 1, meaning the entries at row $i$, column $j$ and row $j$, column $i$ in the weight matrix are set to 1. The second case is when $ij$ is a local edge as defined in this study; its weight is also set to 1. The third case is when the shortest path length of $ij$ is less than the sum of the shortest path lengths from $i$ to the boundary and from $j$ to the boundary; in this case, the weight remains 1. The fourth case is when none of the above three conditions are met, but the shortest path length of $ij$ is less than a threshold $\tau$; then the weight is set to $w_0$. If none of these four cases are satisfied, the weight is set to 0. Based on experience, setting $w_0$ to 0.3 is appropriate. $\tau$ varies across different sites; in practice, choosing $\tau$ involves a trade-off between embedding quality with respect to metric accuracy and topological preservation, and thus requires tuning. Notice that the second requirement for distance function addresses the consistency between topological and metric embedding, which ensures that, after balancing with $\tau$, a satisfactory equilibrium point exists between topological and metric embedding.

The optimization objective is to minimize the weighted squared difference between the embedding distances and the path distances, using the weight matrix defined above. The optimization procedure utilizes the SMACOF algorithm (Scaling by Majorizing a Complicated Function), and is accelerated through Reduced Rank Extrapolation (RRE)\cite{Rosman2010}. Specifically, for each site, five RRE sequences are performed, each consisting of 50 iterations. The final 10 iterations of each sequence are linearly interpolated to generate the starting point for the next sequence. This configuration reliably achieves convergence across all sites.

\subsection{The location choice problems on the human mobility manifolds}\label{subsec3}
\subsubsection{Location choice as a set cover problem}
In this study, the optimized location of a set of generic service facilities comes from the set cover problem of maximizing the coverage of the individual clients\cite{Berman2010}. Firstly, the covered individuals by each spatial grid (node) is obtained from trajectory data, through the set of individuals that stayed in the grid.  With the number of chosen grids as the cost, and with the total number of covered individuals as the utility, we can construct the integer programming problem. 

\begin{equation}
\max_l \|\bigcup_{i \in l}c_i\|, s.t. \|l\|\leq K\label{eq3}
\end{equation}

where $c_i$ is the set of individuals covered by node $i$. $l$ is the chosen set of locations. $K$ is a predefined parameter that constrains the number of locations to be selected, and $\|\bullet\|$ denotes the number of elements in the set. The problem background and parameter settings are consistent with a previous study\cite{Zhang2024}.

\subsubsection{The embedding for the location choice problem and its uniformity test}

As mentioned earlier, our embedding method for the location choice takes into account the characteristics of the scenario. Specifically, we define the distance for location choice $x_{ij}^{loc}$ based on the repetitive coverage of the individual sets covered by two nodes,

\begin{equation}
x_{ij}^{loc}=\frac{\min (\|c_i\|, \|c_j\|)}{\|c_i \cap c_j\|}\label{eq4}
\end{equation}

We empirically verify that this distance function satisfies the triangle inequality. For each triangle in the network, we sort the lengths of the three edges and denote them as $e_1,e_2,e_3$ in descending order. Next, we group these triangles by $e_1$, plot $e_2$ on the horizontal axis, and $e_3$ on the vertical axis. The cases that satisfy the triangular inequality will fall within the region enclosed by the three dashed lines in Extended Data Fig. \ref{edfig1}.

The manifold embedding for location choice is performed using the t-SNE algorithm\cite{VanDerMaaten2008}, leveraging its emphasis on preserving neighborhood relationships. Since optimal locations tend to minimize repetitive coverage and are therefore not in each other's neighborhoods, they are embedded far apart in the resulting manifold. Subsequently, we transform the embedding with the goal of population homogeneity: First, we perform a Voronoi tessellation of the points on the manifold; then, with $\|c_i\|$, we draw a cartogram\cite{Gastner2004} of the Voronoi polygons; and finally, we use the geometric centres of the cartogrammed polygons as the nodes in Fig. \ref{eq3}b.

The uniformity test is as follows\cite{Chen2022}: a certain scale of raster is used to define the manifold boundary, and within the manifold boundary, the ratio between the average nearest neighbour distance, $d_a$, and the average distance from a Poisson distribution, $d_e$, is computed as a measure of the uniformity $R$. The higher the uniformity index, the more uniform the layout is. The reason for extracting the manifold boundaries with a raster is that the point cloud on the manifold does not have an overall clumped shape (this is caused by the fact that the data input only retains the nodes through which the trajectory passes, and which are in fact separated by geographic features, such as mountains and rivers on the surface of the earth). 

For the location choice problem, $R$ of the optimal location nodes on the manifold is 2.27, exceeding the value of 1.04 observed in the geographic space. According to the uniformity reference table\cite{Chen2022}, a value of 2.27 corresponds to a triangular dot-matrix distribution, whereas 1.04 corresponds to a completely randomized distribution.

\subsection{The propagation modelling on the human mobility manifolds}\label{subsec4}

\subsubsection{The embedding for the propagation modelling}

We weightedly fuse the spatiotemporal coexistence interactions (as an expression of contact propagation\cite{WHO2023}) with the matrix characterizing the adjacency between nodes (the introduction of the adjacency matrix can be seen as a hybridization of the two mechanisms of skip transmission and surface diffusion\cite{Taylor2015}) to obtain the distance function for propagation modelling:

\begin{equation}
x_{ij}^{prop}=1-\ln (\frac{\chi_{ij}+b \delta_{ij}}{\max_{kl} (\chi_{kl}+b \delta_{kl})}+\epsilon_0)\label{eq5}
\end{equation}

where $x_{ij}^{prop}$ is the distance for propagation modelling. $\chi_{ij}$ is the spatiotemporal coexistence interaction, namely the number of individuals residing in $j$ who previously spatiotemporal coexisted with any individual residing in $i$. $\delta_{ij}=1$ if and only if $i$ and $j$ form a four-way neighbouring. $b$ is a parameter that balances the two propagation mechanisms, and need to be estimated based on how well the defined distance fits the observed infection patterns. The logarithmic form of the distance function is adopted from a previous study on effective distance\cite{Brockmann2013}, and $\epsilon_0$ is introduced to avoid taking the logarithm of zero, with $\epsilon_0=10^{-16}$ in practice. Since all transmissions in this case originate from the central node, any nonzero $\chi$ must have the propagation center at one end; and given that $\delta$ represents adjacency, the triangular inequality is satisfied. The manifold embedding in the propagation modelling problem is performed using the ISOMAP algorithm\cite{Tenenbaum2000}, which leverages its isometric property.

\subsubsection{Isotropic test for the propagation on the human mobility manifolds}

The isotropy test is based on a linear regression between the number of infections and the manifold distance to the propagation center. Starting from the propagation center, the manifold is divided into eight sectors with equal angular width. We analyzed the 95\% confidence intervals of the regression intercepts and slopes for each of the eight sectors individually, as well as for the combined regression across the whole manifold. If the confidence intervals of the individual sectors overlap with that of the overall regression, isotropy is considered to hold. A similar analysis is conducted in the geographic space, with geographic distance replacing manifold distance (Extended Data Fig. \ref{edfig2}). 

\backmatter

\bmhead{Supplementary information}

\bmhead{Acknowledgements}

This work was supported by the National Natural Science Foundation of China (grant number 42371236) and National Key Research and Development Plan of China (grant number 2022YFC3800803). H.J. thanks Guodong Xue and Prof. Lei Dong for discussions and comments.

\begin{itemize}
\item Funding
\item Conflict of interest/Competing interests (check journal-specific guidelines for which heading to use)
\item Ethics approval and consent to participate
\item Consent for publication
\item Data availability: Datasets for UK, Japan, São Paulo, and Italy are freely available from public repositories\cite{Zhong2024,Yabe2024,Flores-Garrido2024,Pepe2020}. Raw mobility data for Shenzhen are not publicly available to preserve privacy. The distance matrix inverted from human mobility flows for Shenzhen to reproduce the findings of this study can be requested from the corresponding authors. 
\item Materials availability
\item Code availability: The code to replicate this research can be found at https://github.com/jianghezhishi/Geography-as-Manifolds.
\item Author contribution
\end{itemize}

\noindent
If any of the sections are not relevant to your manuscript, please include the heading and write `Not applicable' for that section. 

\bigskip
\begin{flushleft}%
Editorial Policies for:

\bigskip\noindent
Springer journals and proceedings: \url{https://www.springer.com/gp/editorial-policies}

\bigskip\noindent
Nature Portfolio journals: \url{https://www.nature.com/nature-research/editorial-policies}

\bigskip\noindent
\textit{Scientific Reports}: \url{https://www.nature.com/srep/journal-policies/editorial-policies}

\bigskip\noindent
BMC journals: \url{https://www.biomedcentral.com/getpublished/editorial-policies}
\end{flushleft}


\bibliography{sn-bibliography}% common bib file
\newpage
\begin{appendices}

\vspace{2em}
\noindent\textbf{\large Extended Data}
\vspace{1em}
\renewcommand{\thefigure}{ED\arabic{figure}}

\begin{figure}[H]
\centering
\includegraphics[width=0.9\textwidth]{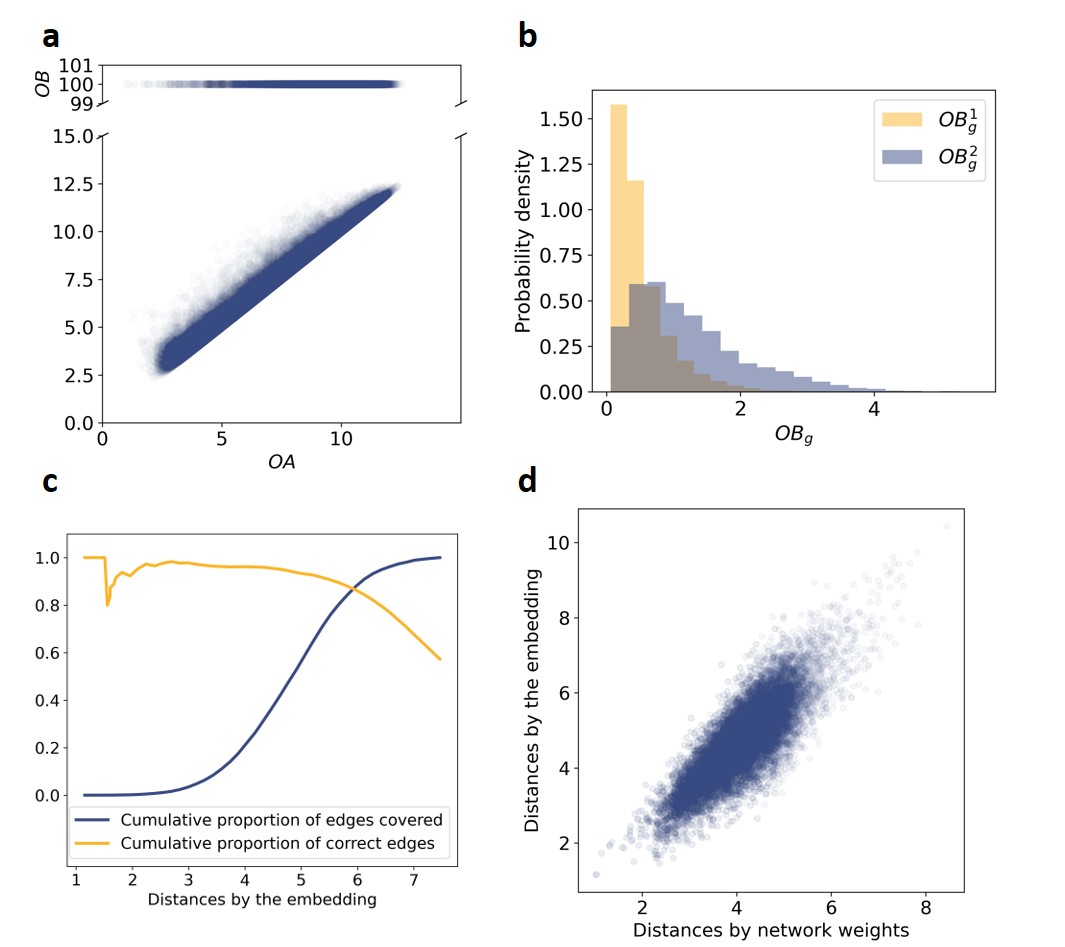}
\caption{\textbf{Human mobility manifolds in England at finer spatial granularity.} Empirical and embedded manifolds of human mobility in England, constructed with one finer level of H3 hexagonal grids (4,840 nodes in total) than in the main analysis. \textbf{a}, Dichotomy analysis for the existence of localities. \textbf{b}, Empirical evidence on mapping between network and geographic localities. \textbf{c}, Topological quality of the embedding, with intersection point at 0.8684. \textbf{d}, Metric quality of the embedding, with Pearson correlation coefficient $r = 0.8259$.}\label{edfig0}
\end{figure}

\begin{figure}[H]
\centering
\includegraphics[width=0.9\textwidth]{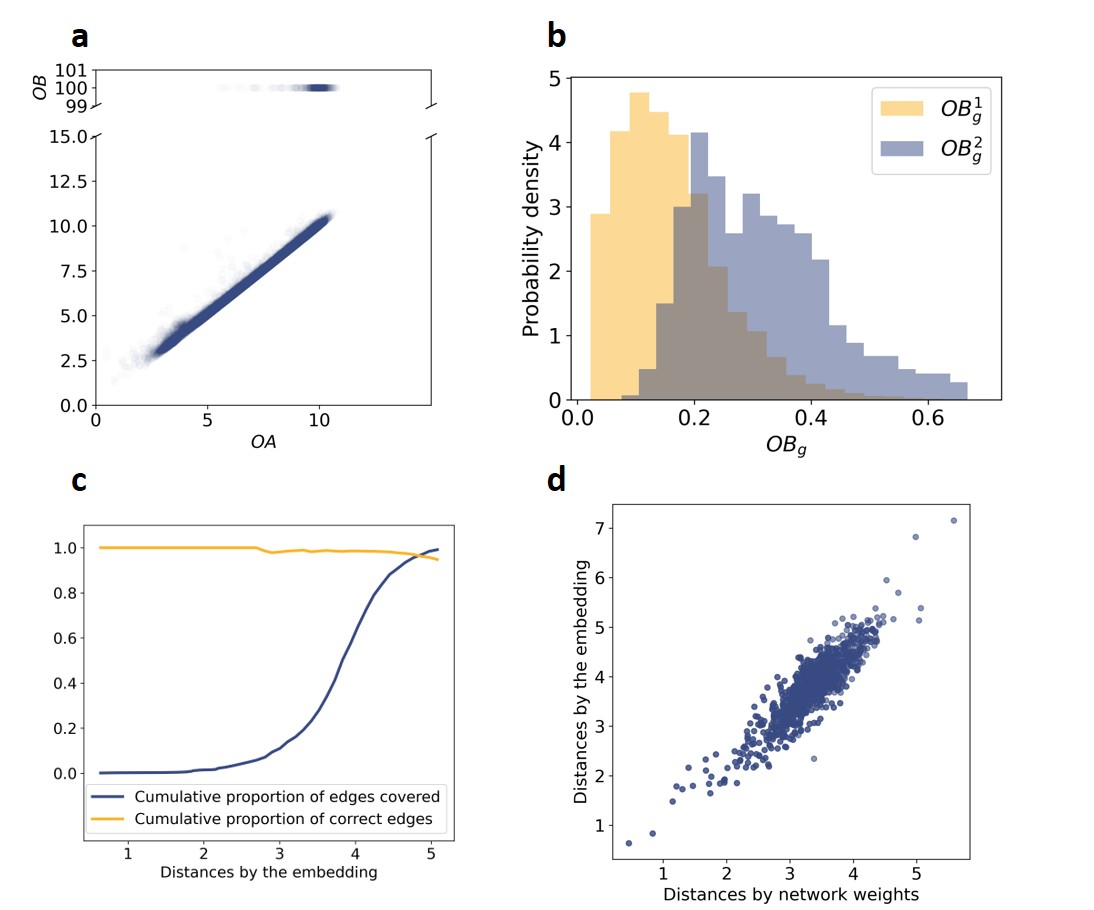}
\caption{\textbf{Human mobility manifolds in London at finer spatial granularity.} Empirical and embedded manifolds of human mobility in London, constructed with H3-level-7 hexagonal grids (342 nodes in total) than in the main analysis. \textbf{a}, Dichotomy analysis for the existence of localities. \textbf{b}, Empirical evidence on mapping between network and geographic localities. \textbf{c}, Topological quality of the embedding, with intersection point at 0.9647. \textbf{d}, Metric quality of the embedding, with Pearson correlation coefficient $r = 0.8912$.}\label{edfig0-1}
\end{figure}

\begin{figure}[H]
\centering
\includegraphics[width=0.9\textwidth]{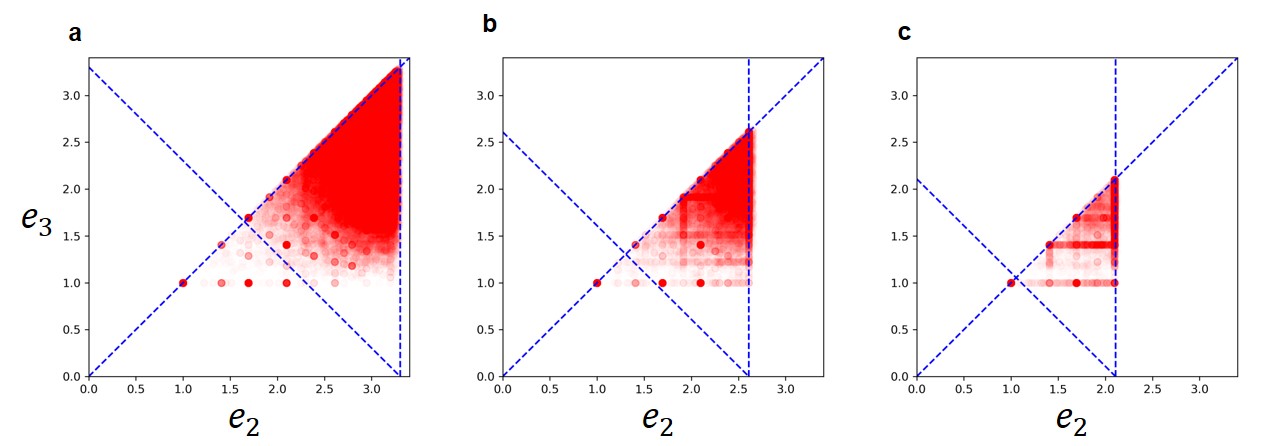}
\caption{\textbf{Empirical analysis of the triangular inequality for the location choice problem.} Panels \textbf{a-c} show the triangles with $1/e_1\in (0.09,0.11)$, $1/e_1\in (0.19,0.21)$ and $e_1\in (0.32,0.34)$, respectively, and the distance function takes the form of Eq. \ref{eq4}. The proportion of combinations outside the region enclosed by the three dashed lines in \textbf{a-c} are 0.9\%, 2.2\%, 4.2\%, respectively. Therefore, the triangle inequality is statistically valid.}\label{edfig1}
\end{figure}

\begin{figure}[H]
\centering
\includegraphics[width=0.9\textwidth]{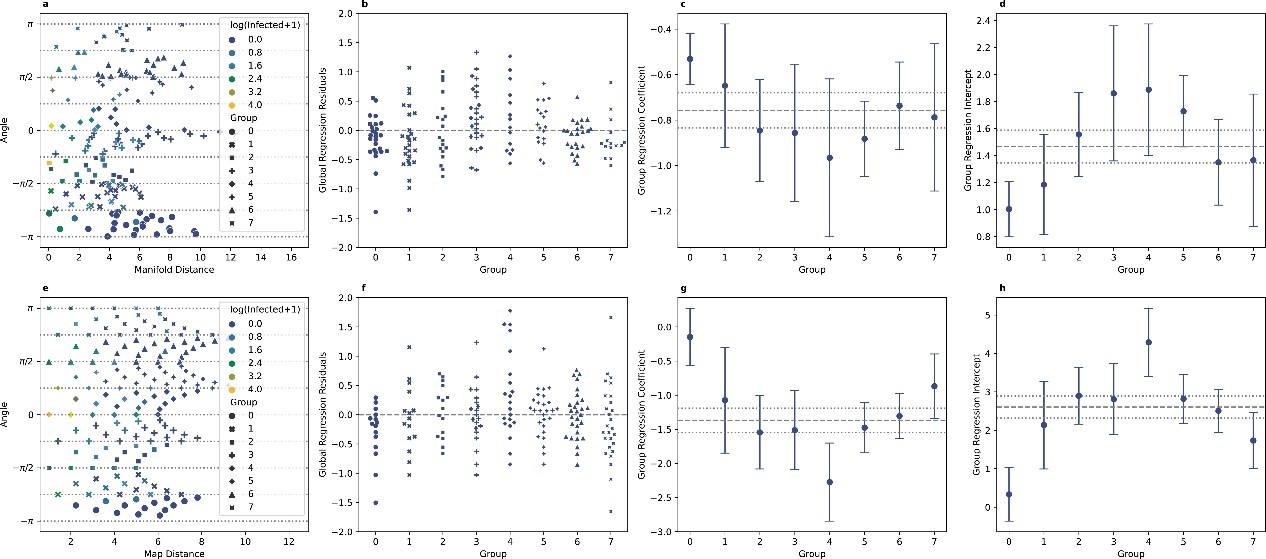}
\caption{\textbf{Isotropic test results of distance decay on the manifold for propagation modelling.}  Panels \textbf{a-d} show the isotropic test of the manifold. It can be seen that the parameter confidence intervals of almost all directions intersect the global confidence intervals, namely between the two dashed lines, indicating isotropy. Panels \textbf{e-h} show the isotropic test for cases on geographic space, and it can clearly be seen that the isotropy is worse than that on the manifold. This phenomenon is evident from Fig. \ref{fig3} \textbf{e,f} in the main text.}\label{edfig2}
\end{figure}

\end{appendices}
\end{document}